*Research article*

# Forecasting the movements of Bitcoin prices: an application of machine learning algorithms

**Hakan Pabuçcu[1,*], Serdar Ongan[2] and Ayse Ongan[3]**

[1] Department of Business Administration, Bayburt University, Bayburt, Turkey
[2] Department of Economics, St. Mary's College of Maryland, Maryland, USA
[3] Fuqua School of Business, Duke University, Durham, USA

* **Correspondence:** Email: hpabuccu@bayburt.edu.tr; Tel: +905545715366.

**Abstract:** Cryptocurrencies, such as Bitcoin, are one of the most controversial and complex technological innovations in today's financial system. This study aims to forecast the movements of Bitcoin prices at a high degree of accuracy. To this aim, four different Machine Learning (ML) algorithms are applied, namely, the Support Vector Machines (*SVM*), the Artificial Neural Network (*ANN*), the Naïve Bayes (*NB*) and the Random Forest (*RF*) besides the logistic regression (LR) as a benchmark model. In order to test these algorithms, besides existing continuous dataset, discrete dataset was also created and used. For the evaluations of algorithm performances, the *F* statistic, accuracy statistic, the Mean Absolute Error (MAE), the Root Mean Square Error (RMSE) and the Root Absolute Error (RAE) metrics were used. The *t* test was used to compare the performances of the SVM, ANN, NB and RF with the performance of the LR. Empirical findings reveal that, while the *RF* has the highest forecasting performance in the continuous dataset, the *NB* has the lowest. On the other hand, while the *ANN* has the highest and the *NB* the lowest performance in the discrete dataset. Furthermore, the discrete dataset improves the overall forecasting performance in all algorithms (models) estimated.

## 1. Introduction

The rapid development of digital currencies during the last decade is one of the most controversial and ambiguous innovations in the modern global economy. Rising technology changes the structure of economies, financial markets and payment methods. The world's financial markets have become more digital than ever before and cashless society is around the corner. Today's technology enables people to create their own money (digital cryptocurrency) and the functions of the central banks, as lenders of last resorts, are discussed and questioned. Bitcoin, as a financial phenomenon, as well as other cryptocurrencies, are in fact data treated like money. Users (called "miners") send and receive these cryptocurrencies (data) electronically from their computers in peer-to-peer network systems to pay for things, if other parties are willing to accept such payments. Market capitalization and the number of miners of 2957 cryptocurrencies reached $221 billion (Bitcoin $147) and 42 million in 2019. The price of Bitcoin has drastically increased from $0.0008 to $10,168 per single coin from being launched in January 2009 to February 2020. Hence, first and foremost, the Bitcoin and other cryptocurrencies have become extremely popular due to increasing number of their users and their huge gains. On the other hand, Bitcoin's and other cryptocurrencies' prices-series, similar to other financial assets-series, exhibit chaotic fluctuations. Because of asymmetric information problems in financial markets, increasing economic-political uncertainties and changing behaviors of miners may make the prices of cryptocurrencies not easily predictable for investors. Cryptocurrencies' forecasting difficulties may well be higher than those of other conventional assets; although they are so popular for investors, very little is known about them, about how they work and how they are created (mined), since they are not physical currencies. Accordingly, accurately forecasting their prices may minimize potential losses-risks for users.

This study aims to forecast the movements of Bitcoin prices at high degree of accuracy. To this end, machine learning (henceforth, ML) algorithms are applied, which do not require strict assumptions like traditional methods, (e.g., regression analysis, discriminant analysis, cluster analysis, etc.). While traditional models use whole data to investigate causal relations, ML algorithms normally split the dataset into training and testing sets. Hence, ML allows computers to "learn" and make predictions. Although both methods try to increase the accuracy by minimizing some loss functions, ML does so using nonlinear algorithms (Butner et al., 2019; Makridakis et al., 2018). This does not mean that ML algorithms always outperform traditional models. However, ML algorithms, specially developed to address specific problems, may provide better forecasts for large datasets. All these make ML algorithms very popular for the scholars to apply.

Many studies empirically compare traditional models and ML algorithms concerning their forecasting performances. In some studies, traditional models outperform ML algorithms, while in others the latter outperform the former. For instance, Jang & Lee (2018) compare the Bayesian neural network and traditional models in forecasting Bitcoin prices. They find that the Bayesian neural network offers higher performance than traditional models. Similarly, McNally et al. (2018) compare the accuracy rates of ML algorithms with auto regressive integrated moving average (ARIMA) model for forecasting Bitcoin prices. They find that ML outperforms the ARIMA model. Rebane et al. (2018) find that the recurrent neural network (RNN) outperforms the ARIMA model in forecasting the prices of cryptocurrencies. Nguyen & Le (2019) apply the ARIMA model and ML algorithms to forecast Bitcoin prices and find that ML algorithms outperform the ARIMA model. Yao et al. (2019) examine the impacts of news articles on Bitcoin prices and find that ML algorithms offer better performance than traditional models. However, Felizardo et al. (2019)

compare the performances of the ARIMA with the RF and the SVM when forecasting Bitcoin prices. They find that the ARIMA model outperforms ML algorithms. Chen et al. (2020) compare traditional models, such as logistic regression and discriminant analysis, with ML algorithms and find that traditional models show better performance in forecasting Bitcoin prices. On the other hand, in some studies, different ML algorithms are compared against each other for their forecasting performances. For instance, Ji et al. (2019) compare the deep neural network (DNN) and long short-term memory (LSTM) for forecasting Bitcoin prices and find that the LSTM slightly outperforms the DNN. Kwon et al. (2019) compare the LSTM and gradient boosting algorithms and find that the LSTM provides a better performance than the gradient boosting algorithm. Furthermore, Miller et al. (2019) use the nonparametric regression method of smoothing splines on 1-minute Bitcoin price data. They find that this method provides better performance than unconditional trading strategies. Lahmiri & Bekiros (2020) use deep learning techniques to forecast the price of the Bitcoin, Digital Cash and Ripple. They find that long-short term memory neural network topologies (LSTM) provides better performance than the generalized regression neural architecture. Huang (2019) use classification tree-based model with 124 technical indicators to investigate cryptocurrency return predictability. They find that this model has strong predictive power. Corbet et al. (2019) find that the variable-length moving average rule performs the best with buy signals for Bitcoin. Atsalakis et al. (2019) use a hybrid Neuro-Fuzzy controller, namely PATSOS, to predict the directional change of the daily price of Bitcoin. They find that performance of the PATSOS system is robust to be used for all cryptocurrencies. Adcock & Gradojevic (2019) find that neural networks provides better performance than various competing models on the prediction of Bitcoin returns. Shu & Zhu (2020) use adaptive multilevel time series detection methodology to predict the bubbles in Bitcoin. They find that this methodology is robust to be used not only on cryptocurrencies but also in financial markets. Balcilar et al. (2017) use non-parametric causality-in-quantiles test to investigate casual relation between trading volume and Bitcoin returns and volatility. They reveal the importance of modelling nonlinearity and accounting on causal relationships. Gyamerah (2019) uses the GARCH models to evaluate the volatility of Bitcoin returns. The author finds that t-GARCH-NIG has the best performance in prediction of the volatilities. Panagiotidis et al. (2018) use the least absolute shrinkage and selection operator (LASSO) framework to examine the effects of factors on Bitcoin returns. The find that gold returns have the most important effects on returns.

This study differs from the studies mentioned in three aspects. First, we apply four different ML algorithms simultaneously to compare their performances. Second, we use nine technical input parameters followed by Armano et al. (2005), Atsalakis & Valavanis (2009), Kara et al. (2011) and Kim (2003). Third, besides the existing continuous dataset, discrete dataset was also created and used. Therefore, all these will enable us to understand which ML algorithm offers higher forecasting performance in continuous and discrete datasets separately and comparatively.

This study is organized as follows. Sections 2 and 3 provide research data preparation and empirical methodology, respectively. Section 4 provides empirical findings obtained from continuous and discrete datasets. Finally, section 5 presents the discussion-conclusion.

## 2. Research data preparation

### 2.1. Continuous data

In this study, for the output, we considered the changes of up and down movements of closing prices from previous days. We coded these as +1 and −1 for ups and downs, respectively. We used the same output for continuous and discrete datasets. Closing, high and low prices were used for computing technical indicators and output as reported in Table 1. Continuous (existing) dataset, between 2008–2019 was normalized for all ML models (n = 1935). We used model validation to compare the performances and significances of the models with benchmark model. The validation dataset is consisting of the number (n = 100) of Bitcoin series between June 2020–October 2020. This validation dataset was divided into 10 sub-datasets with 10 samples for each. For each sample, the movement estimations of each estimated model and accuracy statistics were calculated. The average accuracies of these 10 sub-datasets were bilaterally compared with the LR statistics with *t* test. The accuracy statistics for both continuous and discrete datasets were calculated.

**Table 1.** Selected technical indicators.

| Indicators | Formula |
|---|---|
| Simple 14 days moving average (MA) | $C_t + C_{t-1} + \cdots + C_{t-14}/14$ |
| Simple 14 days weighted moving average (WMA) | $\dfrac{((n) * C_t + (n-1) * C_{t-1} + \cdots + C_{t-14}}{(n + (n-1) + \cdots + 1)}$ |
| Momentum (Mom) | $C_t - C_{t-n}$ |
| Stochastic K% (K%) | $\dfrac{C_t - LL_{t-n}}{HH_{t-n} - LL_{t-n}} * 100$ |
| Stochastic D% (D%) | $\sum_{i=0}^{n-1} K_{t-i}\% / n$ |
| Relative strength index (RSI) | $100 - \dfrac{100}{1 + (\sum_{i=0}^{n-1} Up_{t-i}/n)/\sum_{i=0}^{n-1} Dw_{t-i}/n)}$ |
| Moving average convergence/divergence (MACD) | $MACD(n)_{t-1} + \dfrac{2}{n+1} * (DIFF_t - MACD(n)_{t-1})$ |
| Larry William's R% (LW) | $\dfrac{H_n - C_t}{H_n - L_n} * 100$ |
| Accumulation/distribution oscillator (A/D) | $\dfrac{H_t - C_{t-1}}{H_t - L_t}$ |

Note: Source: (Kara et al., 2011); *n is the number of days accepted as 10 here, $C_t$ *closing price,* $L_t$ *low price ve* $H_t$ *High price.* $DIFF_t: EMA(12)_t - EMA(26)_t$. EMA is exponential moving average, $EMA(k)_t: EMA(k)_{t-1} + \alpha * (C_t - EMA(k)_{t-1})$, $\alpha$ is correction factor. $LL_t$ is the lowest low, $HH_t$ is the highest high for the last t days. $M_t = (H_t + L_t + C_t)/3$, $SM_t = (\sum_{i=0}^{n} M_{t-i+1}/n)$, $D_t = (\sum_{i=1}^{n} |M_{t-i+1} - SM_t|/n)$, $Up_t$ and $Dw_t$ are upward and downward price change at time t respectively…

### 2.2. Discrete data

For creating the discrete dataset, the continuous dataset was converted to −1 or +1 by applying the discretization process. +1 and −1 indicate upward and downward movements, respectively (Patel et al., 2015). This new dataset represents the trend of indicators. The discretization process of each technical input indicator is explained in the following paragraphs.

*The moving average* (MA) and the *weighted moving average* (WMA) represent average price changes over a certain period. The MA, as a most used and simplest indicator, indicates the general direction of the trend. In this paper, 14 days MA and WMA were used for short-term forecasting. If the current Bitcoin price is above the MA or WMA, this means that the trend is upward, and the value is labeled as +1. If the current Bitcoin price is below the MA or WMA, this means that the trend is downward, and the value is labeled as −1. Financial time series like Bitcoin prices exhibits speculative movements. Therefore, long run predictions may provide not accurate results. Hence, in our technical analyses, we used short-term moving averages for 7–14 days. The exponential moving average (EMA) assigns more weight the most recent data. Hence, it smooths the data and thereby provides more importance to the current trend.

*Momentum* (Mom) is an indicator that represents the effect of price changes and presents information of the sustainability of the current trend. If the momentum value is positive, the trend is "upward" and labeled as +1. If the momentum value is negative, the trend is "downward" and labeled as −1. The main problems of determining the momentum boundary line are crisp rises and slumps of time "*t*" for any value, since these changes can affect the momentum boundary line. Momentum is one of the leading indicators which measures velocity of the changes in security prices in a specific period of time. It compares prices of *t* and *t-1* terms.

The *stochastic indicators* K%, D% and LW are clear data trends. A stochastic oscillator, as a one of the most important indicators, determines securities' momentum and identifies the overbought and oversold levels. It utilizes a 0–100 bounded range of values. The LW, developed by Larry Williams, is very similar to the stochastic oscillator and is used in the same way. It compares securities' closing prices and their the high-low ranges over time. If the value of an indicator at time "*t*" is greater than the value at time "*t−1*", the trend is "upward" and labeled as +1 and if the value of an indicator at time "*t*" is lower than the value at time "*t−1*", the trend is "downward" and labeled as −1. A stochastic oscillator tends to vary around some mean price level, since they consider-account an asset's price history as an overbought and oversold signal. It utilizes a 0–100 bounded range of values.

*The Relative strength index* (RSI) charts the speed and scale of directional changes in values. The RSI has different values that determine trend behavior. It measures the speed and magnitude of the changes in recent prices to determine overbought or oversold levels of the prices of the securities. If the value of RSI is lower than 30, it is labeled as +1, higher than 70 is labeled as −1. For values between 30–70, if the value of RSI at time "*t*" is higher than the value at time "*t−1*", the trend is "upward" and labeled as +1, and vice-a-versa.

*Moving average convergence/divergence* (MACD) indicator is related to movements of prices. The MACD, developed by Gerald Appel, shows the relationships between two moving averages of the securities' prices. It is calculated by using the differences of short and long Exponential Moving Averages (EMA). If the MACD increases, then prices increase and if the MACD decreases, then prices decrease. If the value of MACD at time "*t*" is greater than the value at time "*t−1*", the trend is "upward" and labeled as +1, and vice-a-versa.

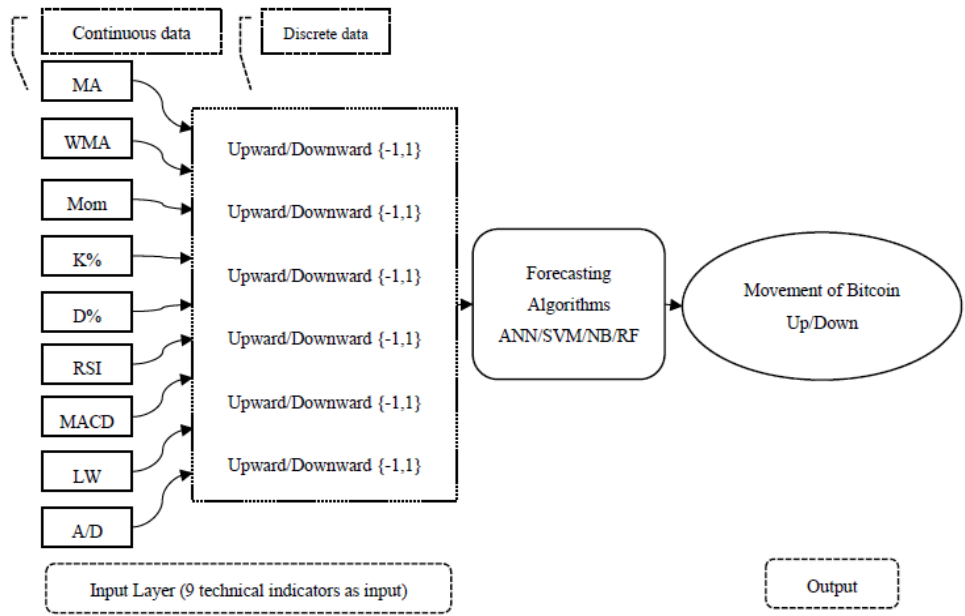

**Figure 1.** Forecasting mechanisms.

## 3. Empirical methodology

Following the calculations of nine technical input parameters, we apply our ML algorithms. The *ANN* (*Artificial Neural Network*) has been commonly used in forecasting price movements. Researchers prefer this algorithm due to its multilayer perceptron (MLP) flexibility (Mallqui & Fernandes, 2019). In this study, tangent sigmoid transfer and logistic transfer functions are used for hidden and output layers, respectively. Threshold was used to predict up and down movements of the Bitcoin prices. Several configurations were tried to determine the best parameter settings for the ANN. Parameter setting levels of the ANN models are reported in Table 2.

**Table 2.** Parameter settings for *ANN*.

| Parameter | Level |
| --- | --- |
| Number of neurons in hidden layer (n) | 5,…,50 |
| Iteration(ep) | 250, 500,…,2000 |
| Momentum constant (mc) | 0.1, 0.2,…,0.9 |
| Learning rate (lr) | 0.1, 0.2, 0.3 |

The *SVM* (*Support Vector Machine*), proposed by Vapnik (1995) is based on a structural risk minimization process by maximizing the margin between negative and positive samples. The *SVM* constructs a hyperplane, which can separate the classes of the real problem (Kara et al., 2011). This is not a stochastic model. It means that it always gives the same results when the same dataset is processed at any given time. In this study, different levels of parameter settings were used to determine the best estimator, as reported in Table 3.

Table 3. Parameter settings for *SVM*.

| Parameter | Level (polynomial) | Level (RBF-Gaussian) |
|---|---|---|
| Kernel function degree ($d$) | 1, 2, 3, 4 | |
| Kernel function Gamma coefficient ($\gamma$) | | 0, 0.1, 0.2,…,5.0 |
| Regularization parameter ($c$) | 1, 10, 100 | 1, 10, 100 |

The *NB* (*Naïve Bayes*) is one of the machine learning classification algorithms based on a conditional probability principle, which is known as Bayes Theorem. Due to the simplicity of its calculation and usage, the *NB* is superior to other machine learning algorithms. This algorithm uses a Bayesian classifier to forecast the probability of samples belonging to a specific class of the given dataset. The *NB* has no other parameter set to construct the forecasting model.

The *RF* (*Random Forest*) is a classification algorithm that is very efficient and offers the opportunity to compare the results with other classification algorithms. ID3 (Quinlan, 1986), C4.5 (Quinlan, 1988) and the CART (Breiman, 1984) are the most powerful and commonly used classification algorithms, known as decision-tree based. The *RF* belongs to an ensemble-learning algorithm based on the idea that a single classifier could not be capable of determining the class of test data. In this study, randomly selected features varying from 3 to 100 and a number of trees varying from 3 to 300 were used to determine the best parameter setting.

The LR (*Logistic Regression*) is a popular technique to model the probability of discrete (i.e., binary or multinomial) outcomes. In this study, this technique, as a benchmark model, was used to compare the performances of machine learning algorithms.

In order to test algorithms mentioned above and compare their performance, *F* statistics are calculated by using the true/false positive (TP-FP) and true/false negative (TN-FN), following the equations below (Patel et al., 2015).

$$Precision_{positive} = \frac{TP}{TP+FP} \tag{1}$$

$$Precision_{negative} = \frac{TN}{TN+FN} \tag{2}$$

$$Recall_{positive} = \frac{TP}{TP+FN} \tag{3}$$

$$Recall_{negative} = \frac{TN}{TN+FP} \tag{4}$$

$$Accuracy = \frac{TP+TN}{TP+FP+TN+FN} \tag{5}$$

$$F = \frac{2*Precision*Recall}{Precision+Recall} \tag{6}$$

Machine learning algorithms do not require stationary tests differently from econometric models. In order to test the performances of selected algorithms, besides *F* statistics, mean absolute error (MAE), root mean square error (RMSE) and root absolute error (RAE) are also used. The continuous dataset was normalized for all models estimated and divided into two parts as training (75%) and testing (25%). Furthermore, a new validation dataset (n = 100) was used for testing the statistical significances of the performance differecences between estimated models and benchmark model.

## 4. Empirical findings

In this section, the estimated model parameters are reported for continuous and discrete datasets, respectively. Descriptive statistics for inputs are reported in Table 4.

**Table 4.** Descriptive statistics for selected indicators.

| Indicator | Minimum | Maximum | Mean | Standard dev. |
| --- | --- | --- | --- | --- |
| MA | 158.407 | 16866.037 | 2501.552 | 3395.42 |
| WMA | 176.498 | 17802.757 | 2507.79 | 3402.815 |
| Mom | −5578 | 8212.55 | 23.237 | 920.879 |
| K% | 0 | 100 | 54.624 | 29.336 |
| D% | 6.337 | 93.153 | 54.343 | 22.571 |
| RSI | 10.954 | 93.491 | 52.549 | 14.209 |
| MACD | −1479.221 | 2520.715 | 12.44 | 292.976 |
| LW | −100 | 0 | -45.376 | 29.336 |
| A/D | −0.879 | 1.521 | 0.399 | 0.185 |

*4.1. Findings of continuous data*

The best parameter combinations are determined by means of experiments for each forecasting algorithm. The estimated best three parameter combinations of the *ANN* models for continuous data are reported in Tables 5–8.

**Table 5.** Best three-parameter combinations for *ANN*.

|   | Learning rate (lr) | Iteration (ep) | momentum constant (mc) | Hidden neuron (n) | Accuracy | MAE | RMSE | RAE |
| --- | --- | --- | --- | --- | --- | --- | --- | --- |
| 1 | 0.3 | 500 | 0.2 | 6 | 0.843 | 0.203 | 0.341 | 0.409 |
| 2 | 0.3 | 500 | 0.2 | 8 | 0.841 | 0.201 | 0.360 | 0.405 |
| 3 | 0.3 | 500 | 0.2 | 7 | 0.835 | 0.201 | 0.349 | 0.404 |

Test results in Table 5 indicate that the accuracy levels and error statistics calculated are within acceptable levels. The best accuracy level is determined as 0.843 for the *ANN*. This means that we will be able to forecast the movements of Bitcoin prices at a high degree of accuracy. After training processes, hidden neurons, momentum constant and learning rates are found as 6, 0.2 and 0.3, respectively. Test results of the best three *SVM* models based on 3 polynomial and gaussian functions are reported in Table 6.

The accuracy statistics are used to determine the best estimated model. Polynomial and radial basis (Gaussian) Kernel functions are used. The best accuracy level is determined as 0.808 with second degree polynomial Kernel, as reported in Table 6. The test results of the *NB* are reported in Table 7.

**Table 6.** Best three-parameter combinations for *SVM*.

|   | Kernel function | d | γ | c | Accuracy | MAE | RMSE | RAE |
|---|---|---|---|---|---|---|---|---|
| 1 | Polynomial | 2 | - | 100 | 0.808 | 0.192 | 0.438 | 0.387 |
| 2 | Polynomial | 1 | - | 30 | 0.804 | 0.196 | 0.443 | 0.395 |
| 3 | Polynomial | 2 | - | 20 | 0.802 | 0.198 | 0.445 | 0.339 |
| 4 | RBF (Gaussian) | - | 0.1 | 20 | 0.733 | 0.266 | 0.516 | 0.538 |
| 5 | RBF (Gaussian) | - | 0.1 | 10 | 0.729 | 0.271 | 0.520 | 0.546 |
| 6 | RBF (Gaussian) | - | 0.1 | 40 | 0.717 | 0.283 | 0.532 | 0.571 |

**Table 7.** *NB* classification parameters.

|   | Accuracy | MAE | RMSE | RAE |
|---|---|---|---|---|
| 1 | 0.626 | 0.368 | 0.572 | 0.743 |
| 2 (Gaussian) | 0.717 | 0.283 | 0.461 | 0.571 |

**Table 8.** Best three-parameter combinations for *RF*.

|   | Feature | Number of tree | Accuracy | MAE | RMSE | RAE |
|---|---|---|---|---|---|---|
| 1 | 3 | 297 | 0.884 | 0.191 | 0.297 | 0.384 |
| 2 | 8 | 251 | 0.882 | 0.180 | 0.293 | 0.362 |
| 3 | 6 | 267 | 0.880 | 0.184 | 0.293 | 0.370 |

**Table 9.** Comparison the best models.

|   | TP | FP | ROC | F-Stat. | Rank |
|---|---|---|---|---|---|
| ANN | 0.843 | 0.149 | 0.910 | 0.843 | 2 |
| SVM | 0.808 | 0.191 | 0.809 | 0.808 | 3 |
| NB | 0.717 | 0.278 | 0.826 | 0.717 | 4 |
| RF | 0.884 | 0.118 | 0.949 | 0.884 | 1 |
| LR | 0.781 | 0.832 | 0.828 | 0.562 | (Benchmark) |

Test results indicate that the accuracy level is determined to be 0.717 for Gaussian *NB* classifiers as the best forecasting algorithm. Test results for the *RF* model are reported in Table 8.

In Table 8, a number of trees are selected as parameter for the *RF*. It ranges from 50 to 300 during the best parameter selection process and it uses 1 to 10 features to train the trees. The best accuracy level is selected as 0.884 for *RF* with 3 features and 297 trees. Performance comparisons of the models described above are reported in Table 9.

Test results in Table 9 indicate that, while the Gaussian process *NB* model presents the lowest performance at 0.717, the *RF* model has the highest at 0.884 value of *F* statistic. The performance differences of the ANN, RF, SVM and NB algorithms with the LR model are statistically significant and they provide better performances compared to the LR model. The results of *t* tests were reported in Table 15.

## 4.2. Findings of discrete data

The best parameter combinations for discrete dataset were determined by means of experiments for each of the forecasting algorithms using discrete data. The selected best three parameter combinations of all models except *NB* are reported in Tables 10–13.

**Table 10.** Best three-parameter combinations for *ANN*.

|   | Learning rate (lr) | Iteration (ep) | momentum constant (mc) | Hidden neuron (n) | Accuracy | MAE | RMSE | RAE |
|---|---|---|---|---|---|---|---|---|
| 1 | 0.3 | 500 | 0.2 | 20 | 0.9483 | 0.072 | 0.206 | 0.480 |
| 2 | 0.1 | 500 | 0.1 | 20 | 0.9463 | 0.077 | 0.207 | 0.512 |
| 3 | 0.1 | 500 | 0.1 | 20 | 0.9395 | 0.086 | 0.214 | 0.546 |

**Table 11.** Best three parameter combinations for *SVM*.

|   | Kernel function | d | γ | c | Accuracy | MAE | RMSE | RAE |
|---|---|---|---|---|---|---|---|---|
| 1 | Polynomial | 3 | - | 1 | 0.9463 | 0.054 | 0.232 | 0.358 |
| 2 | Polynomial | 3 | - | 2 | 0.9442 | 0.056 | 0.236 | 0.372 |
| 3 | Polynomial | 2 | - | 1 | 0.9421 | 0.058 | 0.240 | 0.386 |
| 4 | RBF (Gaussian) | - | 0.2 | 1 | 0.9483 | 0.052 | 0.227 | 0.344 |
| 5 | RBF (Gaussian) | - | 0.2 | 100 | 0.9463 | 0.054 | 0.232 | 0.358 |
| 6 | RBF (Gaussian) | - | 0.1 | 10 | 0.9442 | 0.056 | 0.236 | 0.372 |

Test results in Table 10 indicate that accuracy levels and error statistics calculated are within acceptable levels. The best accuracy level is determined as 0.948 for the *ANN*. After training processes, hidden neurons, momentum constant and learning rates are found to be 20, 0.2 and 0.3, respectively. Test results for the *SVM* model are reported in Table 11.

The error statistics and accuracy statistic are used to determine the best estimated models. Polynomial and radial basis (Gaussian) Kernel functions are used. The best accuracy level is calculated to be 0.9483 and Gaussian Kernel functions are used with the 0.2 Gamma coefficient, and 1 as a regularization parameter. Test results of the *NB* model are reported in Table 12 below.

**Table 12.** *NB* classification parameters.

|   | Accuracy | MAE | RMSE | RAE |
|---|---|---|---|---|
| 1 | 0.8822 | 0.136 | 0.310 | 0.905 |
| 2 (Gaussian) | 0.8822 | 0.134 | 0.309 | 0.905 |

Test results in Table 12 indicate that the best accuracy level for the *NB* classifiers is estimated to be 0.882 with lower MAE and RMSE by fitting the multivariate Bernoulli distribution. Test results of the *RF* model are reported in Table 13 below.

Table 13. Best three-parameter combinations for *RF*.

|   | Feature | Number of tree | Accuracy | MAE | RMSE | RAE |
|---|---|---|---|---|---|---|
| 1 | 10 | 79 | 0.9462 | 0.076 | 0.205 | 0.509 |
| 2 | 8 | 71 | 0.9438 | 0.078 | 0.208 | 0.513 |
| 3 | 10 | 69 | 0.9390 | 0.085 | 0.212 | 0.538 |

A number of trees was selected for the *RF* algorithm. It ranged from 50 to 300 during the best parameter selection process and it used 1 to 10 features to train the trees. The best accuracy level is determined as 0.946 for the *RF* with 10 features and 79 trees. Performance comparisons of these models are reported in Table 14.

Test results in Table 14 indicate that, while the *NB* with the multivariate Bernoulli distribution has the lowest performance at 0.902, the *ANN* presents the highest accuracy at 0.941 value of *F* statistic. For discrete dataset, the ANN, RF, SVM and NB algorithms provided higher performance with higher accuracy and F statistics compared to the LR model as shown in Figure 2. The performance differences are statistically significant for all compared groups as shown in Table 15. Hence, the ANN, RF, SVM and NB algorithms produced better Bitcoin movement predictions compared to benchmark model.

Table 14. Comparison the best models.

|  | TP | FP | ROC | F-Stat. | Rank |
|---|---|---|---|---|---|
| ANN | 0.948 | 0.557 | 0.931 | 0.941 | 1 |
| SVM | 0.948 | 0.610 | 0.669 | 0.938 | 3 |
| NB | 0.882 | 0.167 | 0.901 | 0.902 | 4 |
| RF | 0.946 | 0.557 | 0.923 | 0.939 | 2 |
| LR | 0.858 | 0.873 | 0.681 | 0.854 | (Benchmark) |

Table 15. *t* test results of model comparisons in terms of benchmark.

| Dataset | Model | Mean (Accuracy) | N | Std. Dev. | t |
|---|---|---|---|---|---|
| Continuous | LR | 0.551 | 10 | 0.719 | |
|  | ANN | 0.826 | 10 | 0.017 | −13.658* |
|  | RF | 0.854 | 10 | 0.056 | −9.467* |
|  | SVM | 0.754 | 10 | 0.064 | −11.809* |
|  | NB | 0.657 | 10 | 0.040 | −4.262** |
| Discrete | LR | 0.623 | 10 | 0.032 | |
|  | ANN | 0.850 | 10 | 0.037 | −13.208* |
|  | RF | 0.835 | 10 | 0.012 | −16.582* |
|  | SVM | 0.786 | 10 | 0.053 | −7.398* |
|  | NB | 0.674 | 10 | 0.027 | −2.908* |

Note: *shows the statistical significance at level 0.01 ** shows the statistical significance at level 0.05.

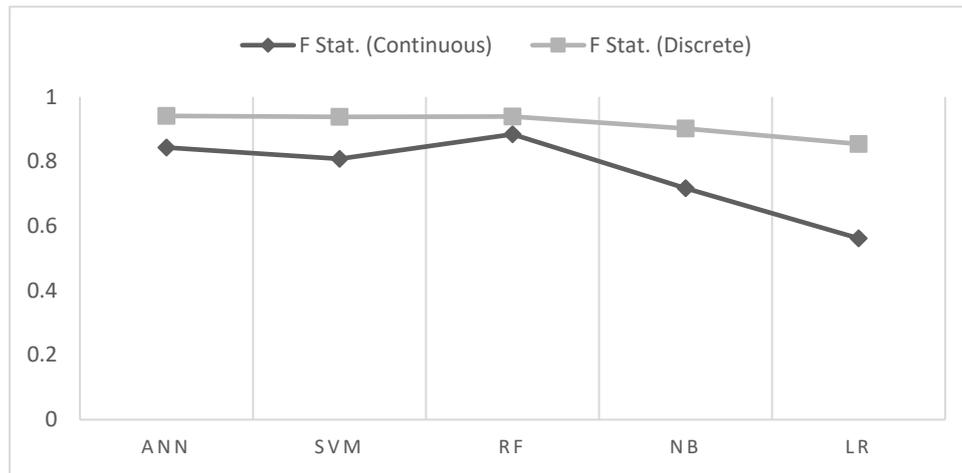

**Figure 2.** F statistics with continuous and discrete data for all models.

## 5. Discussion and conclusion

This study aims to forecast the movements of Bitcoin prices at a high degree of accuracy. To this aim, four different Machine Learning algorithms are applied, namely the Artificial Neural Network, Random Forest, Support Vector Machines, the Naïve Bayes and besides to the logistic regression (LR) as benchmark model. In order to test these algorithms, besides existing continuous dataset, a discrete dataset was also created and used.

Empirical findings reveal that, while the *RF* has the highest forecasting performance, the *NB* has the lowest in continuous dataset. On the other hand, while the *ANN* has the highest performance, the *NB* has the lowest in discrete dataset. Furthermore, discrete dataset improves the overall forecasting performance in all models estimated. The RF has become more popular than the ANN with its ease of use. However, these comparisons potentially can change with new datasets. Furthermore, it should be noted that it will be hard to consider all combinations in a single study. Hence, the performances of the Machine Learning algorithms increase over time.

Each of the nine technical parameters used in this study can also be considered as an estimator. However, these parameters were used after considering their trend characteristics rather than their direct usage as estimators, since this transformation may increase the forecasting performance. This means that with this transformation done in this study, the real-time expert systems may provide advantages to investors to allow for more profitable and safe investments. Furthermore, although it is widely accepted that preprocessing data is not necessary when ML algorithms are used, this study reveals that preprocessing data increases forecasting performances. In this study, algorithms classify Bitcoin prices as up-down. However, instead of only two categories, it is suggested that multi-categories using different algorithms may be used for future forecasts.

This study shows the need for more empirical studies using other techniques to ensure more accurate forecasts for the movements of the Bitcoin price, which exhibit chaotic and nonlinear characteristics (fluctuations) in conditions of increasing economic uncertainties. At this point, besides the nine technical parameters used in this study, some other macroeconomic parameters, such as exchange rate, interest rate, government policy implementations, are proposed for use in these models as new inputs (variables), because all these variables may easily affect the financial markets involving cryptocurrencies.

**Conflict of interest**

The authors declare no conflicts of interest in this paper.